\documentstyle[11pt,newpasp]{article}

\font\ninerm=cmr9    \font\sixrm=cmr5
  
\font\nineit=cmti9  
\font\ninesl=cmsl9
\font\ninei=cmmi9    \font\sixi=cmmi5
\skewchar\ninei='177
\font\ninesy=cmsy9  \font\sixsy=cmsy5
\skewchar\ninesy='60
\font\ninebf=cmbx9  \font\sixbf=cmbx5
\font\nineex=cmex10 scaled 833

\font\ninett=cmtt9
\def\adjustlinespace{\baselineskip=\baselineskip}
\def\ninepoint{\textfont0=\ninerm \scriptfont0=\sixrm 
                \def\rm{\fam0\ninerm}\relax
                \textfont1=\ninei \scriptfont1=\sixi 
                \def\mit{\fam1}\def\oldstyle{\fam1\ninei}\relax
                \textfont2=\ninesy \scriptfont2=\sixsy 
                \def\cal{\fam2}\relax
                \textfont3=\nineex \scriptfont3=\nineex 
                \def\it{\fam\itfam\nineit}\relax
                \textfont\itfam=\nineit
                \def\sl{\fam\slfam\ninesl}\relax
                \textfont\slfam=\ninesl
                \def\bf{\fam\bffam\ninebf}\relax
                \textfont\bffam=\ninebf \scriptfont\bffam=\sixbf
                \def\tt{\fam\ttfam\ninett}\relax
                \textfont\ttfam=\ninett
                \setbox\strutbox=\hbox{\vrule
                     hnine7pt depth2pt width0pt}\baselineskip=9pt
                \adjustlinespace
                \rm}
\font\caps=cmcsc10			   %caps & small caps (IAU etc)
\font\fivebmi=cmmib6
\font\sixbmi=cmmib6	\skewchar\sixbmi='177
\font\ninebmi=cmmib10 at 9pt 	\skewchar\ninebmi='177
\newfam\bmifam
\textfont\bmifam=\ninebmi
\scriptfont\bmifam=\sixbmi
\scriptscriptfont\bmifam=\fivebmi

\def\etal{{\em et~al.\ }}
\def\aa #1 #2 {A\&A, #1, #2}
\def\aas #1 #2 {A\&AS, #1, #2}
\def\acm #1 #2 {ACM-Trans Math Software, #1, #2}
\def\ada #1 #2 {Ann Astrophys, #1, #2}
\def\agabstr #1 #2 {Astr Ges Abstr Ser, #1, #2}
\def\aj #1 #2 {AJ, #1, #2}
\def\anach #1 #2 {Astr Nachr, #1, #2}
\def\apj #1 #2 {ApJ, #1, #2}
\def\apjl #1 #2 {ApJL, #1, #2}
\def\apjs #1 #2 {ApJS, #1, #2}
\def\araa #1 #2 {ARAA, #1, #2}
\def\apss #1 #2 {ApSpaceS, #1, #2}
\def\celmech #1 #2 {Cel Mech, #1, #2}
\def\esom #1 #2 {ESO Messenger, #1, #2}
\def\fundcp #1 #2 {FunCosP, #1, #2}
\def\jcp #1 #2 {J Comp Phys, #1, #2}
\def\jfm #1 #2 {J Fluid Mech, #1, #2}
\def\jmp #1 #2 {J Math Phys, #1, #2}
\def\ma #1 #2 {Mitt Astr Ges, #1, #2}
\def\mn #1 #2 {MNRAS, #1, #2}
\def\nat #1 #2 {Nat, #1, #2}
\def\obs #1 #2 {Observatory, #1, #2}
\def\pasj #1 #2 {PASJ, #1, #2}
\def\pasp #1 #2 {PASP, #1, #2}
\def\phyr #1 #2 {PhysRep, #1, #2}
\def\physd #1 #2 {Physica D, #1, #2}
\def\rpp #1 #2 {RepProgPhys, #1, #2}
\def\ssr #1 #2 {Sp Sci Rev, #1, #2}
\def\iau127#1{in de Zeeuw P.T. ed, Structure and Dynamics of 
     Elliptical Galaxies, IAU Symp.~No.~127. Reidel, Dordrecht, p.~#1}

\def\inbook#1#2#3#4#5#6{in: #1%
\if#2-%
\else%
, #2%
\fi%
\if#3-%
\else%
, ed.\ #3%
\fi%
\if#5-%
 {\if#4-%
 \else,%
   (#4)%
 \fi}%
\else%
 {\if#4-%
, (#5)%
\else%
, (#5:#4)%
\fi}%
\fi%
\if#6-%
.%
\else%
, #6.%
\fi%
}

\def\spose#1{\hbox to 0pt{#1\hss}}
\def\lta{\mathrel{\spose{\lower 3pt\hbox{$\mathchar"218$}}
     \raise 2.0pt\hbox{$\mathchar"13C$}}}
\def\gta{\mathrel{\spose{\lower 3pt\hbox{$\mathchar"218$}}
     \raise 2.0pt\hbox{$\mathchar"13E$}}}
\def\equal{\! = \!}

\def\ssim{\!\sim\!}

\def\=#1{\overline{#1}}

\def\df{{\caps df}}

\def\kms{{\rm\,km\,s^{-1}}}

\def\pc{{\rm\,pc}}
\def\kpc{{\rm\,kpc}}

\def\msun{{\rm\,M_\odot}}

\def\yr{{\rm\,yr}}

\def\rms{{\caps rms}}

\def\spar{\sigma_{\scriptscriptstyle \Vert}}

\def\mstar{m_\ast }

\begin{document}

\title{Dynamical masses, time-scales, and evolution of star clusters}
\author{Ortwin Gerhard}
\affil{Astronomisches Institut, Universit\"at Basel,\\Venusstr.~7, 
CH-4102 Binningen, Switzerland}

\begin{abstract}
This review discusses (i) dynamical methods for determining the
masses of Galactic and extragalactic star clusters, (ii)
dynamical processes and their time-scales for the evolution of
clusters, including evaporation, mass segregation, core collapse,
tidal shocks, dynamical friction and merging. These processes lead
to significant evolution of globular cluster systems after their
formation.
\end{abstract}

\section{Introduction}
\label{secintro}

The Milky Way and probably all large galaxies contain old globular
cluster populations (see the review by Harris 1991).  These old star
clusters have an approximately log-normal luminosity function, and the
mean cluster luminosity is somewhat brighter than M$_V=-7$ with little
dependence on the host galaxy luminosity. In the Milky Way their
typical mass-to-light ratios are $M/L_V\simeq 2$, and typical total
masses are $\sim 2\times 10^5\msun$ (Pryor \& Meylan 1993). It is
widely assumed that the globular clusters we see today must be the
part of an initially larger population that survived the internal and
external dynamical processes leading to cluster destruction (e.g.,
Ostriker 1988).

One of the exciting results from HST has been the discovery of young
star clusters in starburst and interacting galaxies. Whitmore \&
Schweizer (1995) found many hundreds of young clusters in the Antennae
galaxies.  Young cluster systems have now been discovered in other
interacting and merging galaxies, in barred and starburst galaxies,
and even dwarf starburst galaxies (e.g., ESO 338-IG04, Oestlin, Bergvall
\& Roennback 1998). The luminosity functions of the young clusters are
not log-normal, but seem to be better described by power-laws, about
$\propto L^{-2}$. Carlson \etal (1999) use population synthesis models
to determine the ages of the blue clusters in the young merger remnant
NGC 3597. Based on these models they argue that the difference in the
observed luminosity function when compared to the Galactic globular
clusters cannot simply be an age effect, even if the young clusters
formed with an intrinsic age spread.  Are these young cluster systems
then a good model for what the Milky Way's globular cluster population
could have looked like at birth?

This review gives a brief discussion of dynamical methods to determine
masses of distant and nearby star clusters (Section 2). It then goes
on to describe a number of dynamical processes and their time-scales
which will lead to evolution and potentially destruction of star
clusters over long time-scales.  Finally, the results of some
evolution calculations for globular cluster systems are briefly
summarized (Section 3).

\section{Dynamical Mass Determination for Star Clusters}

In this Section we discuss methods for estimating star cluster
masses from structural and kinematic measurements. Mass estimates
based on stellar population properties are discussed in 
U.~Fritze von Alvensleben's article in these proceedings.

\subsection{Virial Masses}
\label{secvirialmasses}

A simple global mass estimate for a star cluster can be obtained
from the virial theorem. This says that, in equilibrium, the radius
of a stellar system is proportional to $GM/V^2$, where $M$ is the
total mass and $V$ the \rms\ three-dimensional velocity of the stars.
The constant of proportionality generally depends on the stellar density
profile, but Spitzer (1969) showed that if the relation is expressed
in terms of the half--mass radius $r_h$, this dependence is weak
and the constant is approximately 0.4 for realistic cluster profiles.
If we furthermore assume that the cluster is spherical, 
$V^2=3\spar^2$, where $\spar$ is the one-dimensional \rms\ velocity
dispersion along the line-of-sight, and write
$\sigma_{10}=\spar/10\kms$ and $r_5=r_h/5\pc$, then
\begin{equation}\label{virialeq}
  M_V = 7.5 \spar^2 r_h / G = 8.7\times 10^5 \sigma_{10}^2 r_5 \msun.
\end{equation}
When using this formula to estimate star cluster masses from observed
velocity dispersions and radii, a few points should be noted:

(i) Because the virial mass (\ref{virialeq}) is a global estimate, it is
independent of velocity anisotropy. For example, shifting some stars
to radial orbits while keeping the (spherical) potential fixed, will
result in a larger central velocity dispersion but also lead to
reduced velocities in the cluster halo. To maintain virial equilibrium
these changes must add in just such a way that the global $\spar$
remains the same.

(ii) The dynamical evolution of star clusters leads to mass
segregation and the formation of a halo of low-mass stars on
preferentially radial orbits (see \S\ref{secevap} below). For evolved
clusters the measured half-light radius will therefore in general
underestimate the half-mass radius, the observed velocity
dispersion will underestimate the \rms\ velocity dispersion, and
eq.~(\ref{virialeq}) will underestimate the mass. 

(iii) Sometimes only the velocity dispersion for stars in the core is
known, or the velocity dispersion from integrated light within some
aperture.  In these cases, a dynamical model is needed to convert this
to the \rms\ $\spar$. This introduces some uncertainty in the mass
estimate because the derived $\spar$ depends on anisotropy.

With high-resolution spectra and HST photometry virial masses can be
determined for some young `superclusters' seen in starburst
galaxies. Masses for three clusters in M82 are compared by Smith \&
Gallagher in these proceedings, spanning a range from $3\times
10^5\msun - 2\times 10^6\msun$.

\subsection{Core Masses}
\label{seccoremasses}

Rood \etal (1972) gave a formula that is often used to determine {\it
core} masses. This is based on the dynamics of King models (King 1966;
Binney \& Tremaine 1987) and assumes that the velocity
distribution in the core is isotropic:
\begin{equation}\label{cmasseq}
 \left( {M\over L} \right) = { 9 \sigma_0^2 \over 2\pi G I_0 r_c}.
\end{equation}
Here $\sigma_0$ is the central velocity dispersion, $I_0$ the central
surface brightness and $r_c$ the core radius. The product of the two last
quantities is rather insensitive to errors caused by
seeing. Eq.~(\ref{cmasseq}) is very accurate as long as the assumption
of isotropy is met (Richstone \& Tremaine 1986); but it can
overestimate the mass by a factor $\ssim 2$ if the system is actually
radially anisotropic (Merritt 1988).

Core mass-to-light ratios can only be determined for well-resolved
Galactic star clusters for which the core parameters $r_c$, $I_0$ and
$\sigma_0$ can be estimated. Even with the resolving power of HST
the cores of distant young clusters cannot be resolved.

\subsection{Masses from Model Fitting}
\label{secmodelmasses}

An alternative method of estimating cluster masses is fitting the
photometric and kinematic data with dynamical models. A simple such
scheme was used by Djorgovski \etal (1997) in their study of the M31
globular clusters mass-to-light ratios. They used structural and
photometric parameters for these clusters obtained with HST and
kinematic measurements in a rectangular aperture obtained with Keck
and HIRES. They then estimated an aperture correction from King models
to transform their measurements to central velocity dispersions, and
used a formula analogous to eq.~(\ref{virialeq}) to estimate masses
with the constant again determined from models.

For some Galactic globular clusters large velocity samples are
available and in such cases much more detailed model fitting is
possible (Pryor \etal 1989). The masses of the Galactic globular
clusters referred to in \S\ref{secintro} (Pryor \& Meylan 1993) have
been determined by these techniques. A recent such study is C\^ot\'e
\etal (1995) who investigated the dynamics of the globular cluster NGC
3201, using a CCD surface brightness profile and a sample of 857
measured stellar radial velocities to trace the velocity dispersion
profile to large radii.

In such work the data are fitted by single- or multi-mass King-Michie
models. In the multi-mass models, a power-law mass
function for the cluster stars is typically assumed, and for each mass bin
$m_i$, a distribution function of King-Michie type (Michie 1963,
Binney \& Tremaine 1987) is used:
\begin{equation}
  f_i(E,J)\propto e^{-\beta J^2} \left( e^{-A_i E} -1 \right).
\end{equation}
 Here $E$, $J$ are the specific energy and angular momentum of a star
in the (spherical) star cluster potential, and $\beta$ can be thought
of as specifying an anisotropy radius.  In the core of the cluster,
stars of different masses are assumed to be in equipartition
(\S\ref{secequipart}), so that $A_i\propto m_i$. In the fitting
procedure the free parameters are the radius, velocity and luminosity
scale, the
cluster's concentration parameter, the anisotropy radius, and the
index of the mass function. These parameters are determined from
fitting to the measured surface brightness and velocity dispersion
profile. This leads to a determination of the {\sl mass-to-light ratio
profile} and {\sl anisotropy profile}, rather than just a single $M/L$
as for the previously described techniques. However, the fit is
non-unique in the sense that adding even fairly large numbers of
faint low-mass stars in the halo (expected there from mass-segregation and
evaporation, see \S\ref{secevap} below) have little effect on the observed
profiles. In their study, C\^ot\'e \etal (1995) find a steady rise in
M/L with distance from the cluster center, as expected from dynamical
evolution theory, and a global $M/L_B\simeq M/L_V \simeq 2.0\pm 0.2$.

\subsection{Masses from Proper Motions}
\label{secpmmasses}

For nearby Galactic globular clusters, it is possible to measure
stellar proper motions in addition to radial velocities. Proper motion
measurements give information about the velocity dispersions in two
directions on the sky (radial along projected $R$, and tangential),
and for a spherically symmetric cluster they are therefore in
principle sufficient to determine the velocity ellipsoid as a function
of radius, and thus the mass profile free of assumptions about
anisotropy. The projected proper motion dispersions $\sigma_R(R)$ and
$\sigma_T(R)$ are related to the intrinsic velocity dispersions
$\sigma_r(r)$ and $\sigma_t(r)$ by Abel integral equations and can
thus be inverted (Leonard \& Merritt 1989).  Moreover, from the
inferred $\sigma_r(r)$ and $\sigma_t(r)$ one can predict the
line-of-sight velocity dispersions $\spar(R)$ and compare with
independent radial velocity data. This provides a check on the
modelling and also can be used to determine the cluster distance. In
terms of global velocity dispersions, $\langle\spar^2\rangle
= (\langle\sigma_R^2\rangle + \langle\sigma_T^2\rangle)/2$
for a spherical cluster and the correct distance.

In an early study along these lines Leonard \etal (1992) investigated
radial velocity and proper motion data for the globular cluster M13.
They concluded that the mean anisotropy of this cluster
$\beta=3(\sigma_R^2-\sigma_T^2)/(3\sigma_R^2-\sigma_T^2)\simeq 0.3$
and that the effect of the anisotropy on the mass determination is
$\sim 20\%$. Much more detailed modelling will be possible with the
large proper motion surveys currently in progress.

\subsection{Non-Parametric Cluster Mass Distributions}
\label{secnpmasses}

With large samples of stellar velocities at hand, radial velocities or
proper motions, it is possible to infer the mass distribution of the
cluster without making specific assumptions like King-Michie stellar
distribution functions. This requires solving the Jeans and projection
equations for the intrinsic density and velocity dispersions under
some smoothness constraint, given the data. I do not give the
equations here, but refer to the papers mentioned below.

When the data consist of several hundred radial velocities, some
assumption about the anisotropy is still needed. Gebhardt \& Fisher
(1995) describe such a non-parametric analysis of radial velocity data
for four Galactic globular clusters, assuming isotropy of the stellar
orbits. With a few hundred stellar velocities in each case the results
are still noisy, but indicate radially increasing $M/L$-profiles as
expected. Merritt, Meylan \& Mayor (1997) describe a similar analysis
of the cluster $\omega$ Centauri, assuming that it is oblate and seen
edge-on, and that it is described by a meridionally isotropic
two-integral model.  They find that the mass distribution cannot be
strongly constrained by their data, but appears to be slightly more
extended than the luminosity distribution.

As discussed above, proper motion data result in two independent
velocity dispersions in the plane of the sky, and thus, within a
spherical model, they contain sufficient information to determine the
anisotropy of the stellar orbits. With sufficiently large data sets it
will therefore be possible to model the anisotropy profile {\sl and}
mass distribution of a spherical cluster non-parametrically.

\section{Dynamical Evolution Processes and their Time-Scales} 

\subsection{Relaxation}
\label{secrelax}

On dynamical time-scales large star clusters ($N>>100$) evolve
collisionlessly. That is, for some time after birth they are described
by a quasi-equilibrium phase-space distribution function (\df) which
is a function of the integrals of motion (or of the stellar orbits) in
the mean field potential.  On longer times-scales, however, the
graininess of the distribution of stars becomes important, and the
dynamical evolution is no longer collisionless. Over a relaxation time
two-particle interactions then deflect the cluster stars from the
orbits they would otherwise have followed in the mean gravitational
potential.

In the approximation of a homogeneous distribution of equal-mass
stars, with density $\rho_0$ and isotropic Maxwellian velocity
distribution with dispersion $\sigma_0$, the two-body relaxation
time is (Spitzer \& Hart 1971) 
 \begin{equation}\label{trzeroeq}
 t_{r0} \; = 0.34 \; \frac{\sigma_{0}^3}{G^{2} \, m_{\ast} \, 
              \rho_0 \, \ln \,\Lambda} 
          = 1.8 \times 10^8 \; \sigma_{10}^{3} \, n_{4}^{-1} \,
              m_{\ast,\odot}^{-2} \, (\ln \Lambda)_{10}^{-1} \;\yr. 
 \end{equation}
 Here $n\equal 10^4\,n_4 \pc^{-3}$ is the number density of stars,
$\sigma_0\equal 10\, \sigma_{10} \kms$ the velocity dispersion,
$m_\ast\equal m_{\ast,\odot}\msun$ is the mean mass per star,
$\Lambda$ is of order the number of stars, and $(\ln \Lambda)_{10} =
(\ln \Lambda)/10$. Note that $t_{r,0}$ is inversely proportional to
the stellar phase space density. Central relaxation times in
globular cluster cores evaluated with eq.~(\ref{trzeroeq}) are $\sim
10^7-10^9\yr$.

The relaxation time often varies by large factors between the
central and outer parts of a stellar system. It is then useful to
define a half-mass relaxation time. For a virialized star cluster
this is obtained from eq.~(\ref{trzeroeq}) by replacing $\rho_0$ with the mean
density inside the system's half-mass radius $r_h= 5 r_{5} \pc$
and $\sigma_0^2$ by one third of the \rms\ $V^2$, and then
using the virial theorem to express $V^2$ through $r_h$
and the total cluster mass $M=10^5 M_5 \msun$.  The result is 
(Spitzer \& Hart 1971) 
 \begin{equation} \label{trhalfeq}
 t_{rh} = \frac{0.14\, N}{\ln 0.4 N}\,\left(\frac{r_h^3}{GM}\right)^{1\over 2} 
    \!\!\!\! = \frac{N}{26\,\log 0.4 N} \, t_d 
       = \, 7.2 \times 10^8 \, M_{5}^{1/2} \, r_{5}^{3/2} \,
         m_{\ast,\odot}^{-1}\, (\ln\,\Lambda)_{10}^{-1} \, \yr,
 \end{equation} 
where $N$ is the total number of stars in the system and 
 \begin{equation} \label{tdyneq}
   t_d \equiv r_h/V = 1.58\, (r_h^3/GM)^{1/2}
	= 8.3\times 10^5 r_{5}^{3/2} M_{5}^{-1/2} \yr
 \end{equation}
 is the dynamical time.
 For comparison with the local formula eq.~(\ref{trzeroeq}), the
fiducial values used in eq.~{(\ref{trhalfeq}) correspond to a
one-dimensional virial velocity dispersion $\sigma \simeq 3.4 \kms$
and a mean density $ n \simeq 96 \pc^{-3}$.

On short time-scales cluster evolution is still collisionless, so long
as $t_d \ll t_{rh}$ (requiring $N \gg 100$); for example, during the
violent relaxation at formation. The resulting quasi-equilibrium \df\
subsequently evolves slowly in response to collisions, which will tend
to drive the system towards an isothermal energy distribution. One
aspect of such slow evolution would be a decrease of ellipticity with
dynamical age (Fall \& Frenk 1985). This could be the reason for
the significantly rounder globular clusters in M31 and the Milky Way
as compared with the LMC and SMC clusters (Han \& Ryden 1994).

\subsection{Evaporation and Core Collapse}
\label{secevap}

Collisions between single stars modify the stellar \df\ in two ways.
The rarer process is {\sl ejection}, in which a single close
encounter leads one of the stars to acquire a velocity greater than
the local escape velocity $v_e$ and to escape from the cluster. The
time-scale for this is $t_{ej} \equiv -N/(dN/dt) \simeq 1.1 \times
10^3 \ln 0.4N \, t_{rh}$ $\sim 10^4 \, t_{rh}$ 
 % \begin{equation} 
 % t_{ej} \; = \; \left(- \frac{1}{N} \; \frac{dN}{dt} \right)^{-1} 
 %        \; = \; 1.1 \, \times \, 10^3 \; ln \, 0.4 \, N \, t_{rh} 
 %           \simeq \, 10^4 \ t_{rh} 
 % \end{equation}
(H\'enon 1969). The more important process of {\sl evaporation} 
is caused by the cumulative effect of many weak encounters, which
gradually increase a star's energy until $v \geq v_e$. It is easy
to show that the \rms\ escape velocity of the cluster is just
 %\begin{equation} 
 %< v_{e}^{2} > \frac{1}{M} \; \int \; d^{3}r \, \rho \,
 %(- \, 2 \phi) \; = \; - \frac{4\, W}{M} \; = \; 4 \, <v^2>, 
 %\end{equation} 
 twice its \rms\ virial velocity. Thus, on average, a particle with
$v \geq 2 V = \sqrt{12}\spar$ will escape. 
For a Maxwellian velocity distribution, a fraction $\epsilon \sim
0.74 \%$ of stars have $v \geq 2 V$; these
stars will escape in one dynamical time, after which the
high-velocity tail is repopulated only in $\sim t_{rh}$. Thus one
expects the evaporation time scale of the cluster to be $  \sim
(\epsilon/t_{rh})^{-1} $; detailed calculations (Spitzer \& Thuan
1972) show that 
 \begin{equation} 
   t_{ev} \, \equiv \, -\, N \, (dN/dt)^{-1} 
       \, \simeq \; 300 \; t_{rh}.
 \end{equation} 
Because the evaporation is dominated by weak encounters, escaping
stars leave the cluster with only very small positive energy; thus
the total energy of the remaining cluster is nearly constant, but
must be shared among a shrinking number of stars.  In virial
equilibrium $N^2/r_h \simeq $ const.\ and thus $\rho \propto N/r_h^3
\propto 1/N^5$ and $t_{rh} \propto N r_h^{3/2}/M^{1/2} \propto
N^{7/2}$.  So as the cluster becomes denser, evaporation accelerates
and the system contracts to negligible mass and radius in finite
time. 
 %\begin{equation} 
 %\rho \; \propto \; N/r_h^3 \; \propto \; 1/N^5;  \qquad  
 %  t_{rh} \; \propto \;
 %         \frac{N r_h^{3/2}}{M^{1/2}} \; \propto \; N^{7/2}.
 %\end{equation}

This evaporation model, however, neglects the fact that the
evolution of the stellar cluster is not homologous and that the rate
of evolution is much faster in the dense core than in the system's
outer parts. Stars gaining energy towards evaporation build up an
extended halo where the time scale for further energy gain increases
strongly, so that these stars may not in fact escape during the age
of the cluster. On the other hand, the dense core loses stars to the
halo on the much faster central relaxation time, and may collapse to
very high densities before $M_{\rm tot}$ and $r_h$ can change much. 

This phenomenon of {\sl core collapse} may be understood as a
consequence of the fact that self-gravitating star clusters have
negative specific heat (Lynden-Bell \& Wood 1968): In virial
equilibrium the total energy $E = - T$, where $T$ is the total
kinetic energy, which is proportional to the virial temperature
$T/M=V^2$. As energy is withdrawn from the cluster,
its kinetic energy increases and so does the virial temperature.
Since $r_h \simeq GM^2/(2 \vert E \vert)$, the cluster thereby
contracts. Vice-versa, an energy production mechanism (e.g., from
binary stars) causes the cluster to cool and expand. Now the
dense core of the cluster may be approximately regarded as a
virialized system in thermal contact with the rest of the cluster. 
It is normally hotter than its surroundings and therefore loses
energy to them through stellar encounters. As a result it shrinks
and becomes yet hotter, loses still more energy to the surrounding
stars, and contracts to formally zero radius in finite time. 

For a single-mass star cluster the late stages of core collapse are
self-similar (Lynden-Bell \& Eggleton 1980, Cohn 1980). As the core
radius 
 \begin{equation}
    r_c \equiv 3\sigma_c / \sqrt{4\pi G \rho_c}
 \end{equation}
shrinks, the central density $\rho_c$ and velocity dispersion 
$\sigma_c$ increase and the core mass $M_c$ decreases according to 
 \begin{equation}
   \rho_c   \propto r_{c}^{- 2.23}, \qquad
   \sigma_c \propto (\rho_c r_c^2)^{1/2} \propto r_c^{-0.11}, \qquad
   M_c      \propto \rho_{c} r_c^3,      \propto r_c^{0.77}
 \end{equation}
until the core radius and mass formally shrink to zero at time $t_{cc}$.
Moreover, the density profile of the cluster outside the collapsing 
core has the same exponent: $\rho \propto r^{-2.23}$ for 
$r_c(t) \ll r \ll r_c(0)$.

The time-scale for core collapse is proportional to the central
relaxation time; for a single mass cluster it is $t_{cc} \simeq 330
\, t_{rc}$ once the collapse is in the self-similar phase (Cohn 1980,
Heggie \& Stevenson 1988). The {\sl total} time until core collapse
in Cohn's (1980) model is $\sim 16$ half-mass relaxation times or
$\sim 60$ initial $t_{rc}$. As Goodman (1993) has emphasized,
the former number depends on the mass distribution of the cluster,
and $t_{cc}/t_{rh}$ will be less than 16 for clusters more centrally
concentrated than Plummer models. By noting that $r_c^{-1} dr_c/dt
\propto t_{rc}^{-1} \propto \rho_c/\sigma_c^{3} \propto
r_c^{-1.89}$, one can solve for the asymptotic time-dependence of
the collapse: 
 \begin{equation}
    r_c   \propto (t_{cc} - t)^{0.53}, \quad
   \rho_c \propto (t_{cc}  - t)^{-1.18}, \quad
 \sigma_c \propto (t_{cc}  - t)^{-0.06}, \quad
   M_c    \propto (t_{cc}  - t)^{0.41}.
 \end{equation}

In summary, the collapse of a single mass cluster occurs in two 
stages (Cohn 1980). The longer part of the evolution is an evaporative 
phase, during which stellar collisions simultaneously populate a halo 
and make the core shrink and become denser. Only towards the end 
does the evolution accelerate and enter the gravothermal instability
phase of self-similar collapse.

\subsection{Equipartition, Mass Segregation, and Multi-Mass Core Collapse}
\label{secequipart}

When the cluster contains different stellar masses, energy can flow
not only from the core to the halo, but also between stars of
different masses. Stellar collisions drive the system towards
equipartition of energy $m_i \langle v_i^2 \rangle$ = const. As
eq.~(\ref{trzeroeq}) shows, relaxation proceeds faster for more massive
stars.  The equipartition time-scale measures the rate at which a
group of heavy stars with masses $m_2$ loses energy to lighter stars
of mass $m_1$ (Spitzer 1969):
 \begin{equation} \label{eqiparteq}
    t_{eq} = \frac{ (\langle v_1^2 \rangle + \langle v_2^2 
 						\rangle)^{3/2}}
                  {8(6\pi)^{1/2}G^2 m_1 m_2 n_1 \ln\Lambda}
           = 1.2\; \frac{m_1}{m_2} \; t_{r0}(m_1)
 \end{equation}
where we have assumed equal temperatures $\langle v_1^2 \rangle =
\langle v_2^2 \rangle$ and used eq.~(\ref{trzeroeq}). Initially,
$\langle v_i^2 \rangle$ is independent of stellar mass; thus the
massive stars lose kinetic energy and sink to the center, while
lighter stars gain kinetic energy in collisions and move outwards, a
process called {\sl mass segregation}. Moreover, eq.~(\ref{eqiparteq})
shows that mass segregation of the massive stars occurs before
relaxation of the cluster as a whole becomes significant.

However, equipartition may never be reached. A simple case
considered by Spitzer (1969) is one with two mass groups such that
the heavy stars are much more massive than the light stars, $m_2 \gg
m_1$, but the total mass in the cluster core is dominated by the
light stars: $M_{2} \ll \rho_1 r_{c1}^{3}$. In this case
equipartition causes the formation of a small subsystem of heavy
stars ($M_2$, $m_2$) in the core of the distribution of lighter
stars. Applying the virial theorem to the subsystem of heavy stars
gives 
\begin{equation} \label{subvirialeq}
  \langle v_2^2 \rangle \; \simeq \; \frac{0.4 \, GM_2}{r_{h2}} \; +
                    \frac{4 \pi\, G\,\rho_{c1}}{3} \; k^2 r^2_{h2}, 
\end{equation} 
where the first term describes the self-energy of the subsystem $M_2$
and the second term its interaction with the system of light stars
($k$ is a constant of order unity). Spitzer (1969) noticed that the
right-hand-side of eq.~(\ref{subvirialeq}) has a minimum when regarded
as a function of $r_{h2}$. An equilibrium can therefore exist only if
$\langle v_2^2 \rangle$ is greater than this minimum, that is,
assuming equipartition, if
\begin{equation} 
  \frac{M_2}{\rho_{1} r_{c1}^{3}} \; \leq \; 4.0\, k^{-1}
      \; \left( \frac{m_1}{m_2}\right)^{3/2}.
\end{equation}
 In other words, if its mass is too large, the subsystem of heavy
stars remains a dynamically independent stellar system with mean
square velocity greater than the equipartition value. It continues
to lose energy to the lighter stars, becoming denser and hotter,
and evolving away from equipartition all the time ({\sl mass
stratification instability}). Fokker-Planck calculations (Inagaki \&
Wiyanto 1984, Cohn 1985) show that in the end the subsystem of heavy
stars core collapses independently from the cluster of light
particles, just like a single mass system. 

The evolution to core collapse with a spectrum of stellar masses has
been considered by Inagaki \& Saslaw (1985) and Chernoff \& Weinberg
(1990). The detailed evolution occurs in several phases: First,
collisions trying to establish equipartition of energy lead to mass
segregation and the formation of a heavy mass core. Then this core
undergoes the gravothermal instability, i.e., contracts while
remaining hotter than the mean temperature of the system and
conducting energy outwards. This collapse accelerates towards core
collapse, and finally goes over into a single-component collapse
which reaches formally infinite central density.  The time scale for
this multi-mass core collapse evolution is faster than that for core
collapse in any single component cluster, typically a factor of a
few faster than for a cluster composed of the heaviest mass alone. 

Deep in collapse, the density slopes of all mass groups $m_k$ are 
characterized by separate power laws in the region where the
heaviest component dominates the potential, such that
approximately (Cohn 1985, Chernoff \& Weinberg 1990)
\begin{equation}
  d\ln\rho_k / d\ln r \simeq -1.89 \,(m_k/m_u) - 0.35,
\end{equation}
where $m_u$ is the mass of the heaviest species in the cluster.
The overall mass profile is then not self-similar.

A multi-mass core collapse, however, may be strongly influenced by
the stellar evolution of the more massive stars. This has two main
effects: First, the mass loss from massive stars through winds may
lead to an overall mass loss from the cluster, and thus cause an
adiabatic expansion. Secondly, the finite stellar life-time $t_{MS}$
limits the time $t_{cc}$ during which they can core collapse, such
that $t_{cc}(\mstar) \, \lta \, t_{MS}(\mstar)$. 
Both effects greatly increase the overall core collapse times; 
compared to a system of point masses within the range
$(0.4-15) m_\odot$, Weinberg \& Chernoff (1989) find an increase by
about a factor of $30-60$ in their globular cluster models,
including the expansion effects.

A reasonable approximation for the stellar lifetime of all but the
most massive stars is $t_{MS} \simeq 9 \cdot 10^9
(\mstar/m_{\odot})^{-2.6} \yr $. Thus if the most massive stars leave
black hole remnants of $3\msun$, these together with tight binaries
will dominate the evolution after $5\cdot10^8\,\yr$, while if
the most massive remnants are $1.4 \msun$ neutron stars, they and
the binaries will dominate after $4\cdot 10^9\,\yr$. The latter
time scale approaches the time expected for core collapse in typical
Milky Way globulars.

\subsection{Reversing core collapse}
\label{secrevcc}

A number of energy source mechanisms can stop core collapse (e.g.,
Goodman 1993): (i) Processes that generate kinetic energy in the core
directly, such as binary stars transferring energy to the field stars
in collisions. (ii) Mass loss processes that heat the core indirectly
above its virial temperature, including: normal stellar evolution,
accelerated stellar evolution by the formation of massive stars in
mergers, and ejection of stars through binaries. In all cases the net
result is adiabatic expansion and cooling of the core.

Only hard binaries contribute to field star heating. Binaries are
hard if their binding energy $E_b = - Gm_1 m_2 / a$ (with $a$ their
semi-major axis) exceeds the mean kinetic energy, $E_b > 3 \mstar
\sigma^2$; those with $E_b < 3 \mstar \sigma^2$ are called soft
binaries. Heggie's law (Heggie 1975, Hut 1983) states that, on
average, hard binaries get harder by collisions with field stars,
and soft binaries get softer. Essentially, the orbital velocity of a
hard binary is on average greater than the velocity of an incoming
field star, and the tendency towards energy equipartition therefore
results in a net transfer of energy to the field star. The opposite
is true for soft binaries, which gain net energy and eventually
dissolve. The binary behaves like a mini-system with negative
specific heat: as energy is withdrawn from it, the orbit shrinks,
the orbital velocity increases, and the binary hardens. When the
binary becomes sufficiently hard, the typical recoil from a 
collision with a single star becomes large, and the binary will
eventually be kicked out of the cluster. Just like in the Sun, the
binaries providing the nuclear energy source will eventually be
'burned'.

Binaries can be formed by a close gravitational interaction of three
stars (`three-body binaries'), by dissipational tidal capture, or at
the time of star formation (`primordial binaries').  To be effective
in reversing core collapse, binaries must have orbital semi-major axes
\begin{equation}
  a < Gm_\ast/3 \sigma^2 \, = \, 3 \,
               \sigma_{10}^{-2} \, m_{\ast,\odot} \, {\rm AU}.
\end{equation}

The formation of a hard three-body binary requires a close encounter
between two stars ($\delta v \simeq v$) with a third star in the
immediate vicinity, such that one of the three stars acquires 
additional energy, leaving the other two as a bound pair. Thus the 
time-scale is (Goodman 1984, Binney \& Tremaine 1987)
\begin{equation}
  t_3 \, \simeq \, (np^{2}\,v)^{-1} \, (np^{3})^{-1} \, 
         \simeq \frac{\sigma^9}{n^2 \; (Gm_\ast)^5} 
         \simeq \, N_c^2 \, \ln N_c \, t_{r0},
\end{equation}
where $p\simeq Gm_\ast / v^2$, $v\!=\! O(\sigma)$ because low relative
velocities dominate, and we have used eq.~(\ref{trzeroeq}) to express Goodman's
result in terms of the central relaxation time and the total
number of stars in the core, $N_c$. This implies that about $1 / N_c
\ln N_c$ three-body binaries form per central relaxation time. In
other words, three-body binaries become important if the final core
collapse is driven by fewer than 100 of the largest mass stars.

\subsection{Tidal field and tidal shocks}
\label{sectidalshocks}

A steady tidal field lowers the energy threshold beyond which stars
are no longer bound to the cluster. It thus increases the mass loss
rates from evaporation, both because the fraction of stars in the
velocity distribution that escape in a dynamical time increases, and
because the decreasing number of stars in the cluster leads to shorter
relaxation times. The Quintuplet and Arches young clusters (Figer
\etal 1999) in the inner Galactic bulge are two clusters for which
these tidal effects are very important (Kim \etal 1999).

In reality, the tidal field is not stationary in the frame of the
cluster stars. This complicates the escape process, but more
importantly it leads to a new dynamical process in cluster evolution,
referred to as gravitational shocking (Ostriker, Spitzer \& Chevalier
1972).  The tidal field acting on the cluster may suddenly increase in
strength when the cluster passes through the disk of its host galaxy,
or when it comes close to the high-density inner bulge near
peri-galacticon of its orbit. In both cases, the perburbations to the
stellar orbits caused by the tidal shock lead to an effective energy
input in the cluster which makes the cluster less bound and
accelerates mass loss from internal processes.

A detailed recent discussion of this process is given by Kundi\'c \&
Ostriker (1995) and Gnedin, Lee \& Ostriker (1999). For stars in the
outer parts of the cluster, the tidal perturbation can be approximated
as impulsive because of the short time-scale of passage through the
galactic disk. In the cluster's central parts, on the other hand, the
stellar orbital time-scales are short and adiabatic invariance reduces
the effects of the perturbation.  Traditionally these effects of the
tidal shock were described by a first-order term $\langle(\Delta
E)_{ts}\rangle$, which denotes the net energy gain averaged over
stellar orbits at a given position in the cluster. Kundi\'c \&
Ostriker (1995) noticed that the second-order term $\langle(\Delta
E)^2_{ts}\rangle$ is typically even more important and competes with
two-body relaxation near the half-mass radius in driving evolution of
the cluster's internal structure. This can speed up core collapse by
a factor of three (Gnedin, Lee \& Ostriker 1999). Cluster
destruction is accelerated; recent modelling of the evolution of the
Milky Way's globular cluster system shows that the typical time to
destruction becomes comparable to the typical age of the Galactic
globulars (Gnedin \& Ostriker 1997).

\subsection{Dynamical friction and merging}
\label{secdynfrict}

As already noted by Tremaine, Ostriker \& Spitzer (1975), massive star
clusters experience dynamical friction against field stars as they
move along their orbits through the host galaxy. Because of the
frictional drag the cluster loses orbital energy and spirals into the
galaxy center, where the tidal field becomes ever stronger and will
eventually dissolve the cluster.

The time-scale for dynamical friction for a cluster on a circular
orbit at initial radius $r_i = 2 r_{i,2} \kpc$ in a singular
isothermal sphere with circular velocity
$v_c=250\, v_{250} \kms$ is (Chandrasekhar 1943, Binney \& Tremaine 1987)
\begin{equation}
  t_{\rm df}={1.17 r_i^2 v_c \over \ln\Lambda\, GM} = 2.64\times 10^{11}
	   \,r_{i,2}^2\, v_{250}\, M_5^{-1} (\ln\,\Lambda)_{10}^{-1}\yr
\end{equation}
where $M_5$ is again the cluster mass in units of $10^5\msun$. The
friction time-scale thus scales with the square of the cluster's
initial radius in the potential, and is inversely proportional to its
mass. It is the inner, most massive clusters which are affected
first.

If we continue to model the inner parts of the host galaxy as an
isothermal sphere with $M_G(r)=v_c^2r/G$ and use the virial theorem
[eq.~(\ref{virialeq})] for the cluster, we can determine the radius at
which the incoming cluster will dissolve as
\begin{equation}
 r_{\rm dis} \equiv r_h\left(M_G(r_{\rm dis})\over M\right)^{1\over 3}
         = {r_h v_c \over \sqrt{7.5}\, \spar}
	 = 46\, r_5\, \sigma_{10}^{-1} v_{250} \pc.
\end{equation}
Young clusters formed in the high-density regions of starburst galaxies
would thus contribute to the build-up of the nuclear bulge after being 
dragged inwards by dynamical friction and tidally shredded by the tidal
field.

In some circumstances it may be possible that several young clusters
are born close enough to eachother to tidally interact and even merge.
To quantify this we use an approximate merging criterion fitted by
Aarseth \& Fall (1980) to the results of N-body merger
simulations. For the escape velocity of the clusters at pericenter $p$
of their relative orbit we take an approximate expression assuming two
overlapping Plummer spheres, $v_e^2(p)=27.6\spar^2/(1+p^2/1.2 r_h^2)^{1/2}$
(see also the discussion in Gerhard \& Fall 1983). Then the criterion
for merging becomes
\begin{equation}
\left( p \over 4r_h\right)^2 + \left( v_p\over 6\spar \right)^2
	\left(1+{p^2\over 1.2 r_h^2}\right)^{1\over2} \lta 1
\end{equation}
where $v_p$ is the relative velocity at pericenter. Here we have used
the virial equation~(\ref{virialeq}), and $\spar$ is again the
one-dimensional \rms\ velocity dispersion of the cluster. For head-on
collisions, this formula predicts merging for $v_p\lta 6\spar =
60\,\sigma_{10} \kms$ (slightly more than $\sqrt{2}$ times the \rms\
escape velocity from each cluster), or $\Delta v=\sqrt{36-27.6}\, \spar
\simeq 3\spar \simeq 30\, \sigma_{10} \kms$ for their relative velocity
at large separations. It also shows that merging requires the two
clusters to come within several half-mass radii of eachother for
merging to occur, at correspondingly smaller approach velocities.  The
most likely situation for this to happen would be when two clusters
are born from the same giant molecular cloud complex.

\subsection{Evolution of globular cluster systems}
\label{secevolution}

The evolution of globular cluster systems has recently been modelled
in a number of studies, among others by Gnedin \& Ostriker (1997), Murali
\& Weinberg (1997), Baumgardt (1998) and Vesperini (1998). These models
combine assumptions about the initial cluster mass function and
cluster locations with evolutionary models for individual clusters.
In the models the various processes described above are considered,
and treated in some studies by parametrized mass loss rates or
analytic approximations to the results of N-body simulations, in others
as diffusion terms in Fokker-Planck models. Some of the main results
of these studies are:

(i) Globular cluster systems in galaxies evolve significantly.
In the Milky Way the typical cluster destruction time
is of order the age of the system, and about half of the present
globulars will be destroyed in the next Hubble time.

(ii) Clusters in the inner regions of their host galaxy are
disrupted most rapidly. Similarly, clusters on eccentric orbits
are preferentially destroyed over clusters on tangential orbits.

(iii) Low-mass and high-concentration clusters are disrupted by
evaporation, loosely bound clusters and those on central or eccentric
orbits by tides, and massive inner clusters by dynamical friction
and tides.

(iv) Low-mass clusters are destroyed most efficiently and initial
power-law mass distributions tend to become transformed towards
approximately log-normal mass distributions.

\end{document}